\documentclass[twocolumn,amsmath,amssymb,floatfix]{revtex4}

\usepackage{graphicx}% Include figure files
\usepackage{dcolumn}% Align table columns on decimal point
\usepackage{bm}% bold math

\graphicspath{{Figures/}} \setkeys{Gin}{width=\linewidth}

\begin{document}

\title{Hole transport in p-type GaAs quantum dots and point contacts}

\author{B. Grbi\'{c}$^{*}$, R. Leturcq$^{*}$, T. Ihn$^{*}$,
K. Ensslin$^{*}$, D. Reuter $^{+}$, and A. D. Wieck$^{+}$}

\affiliation{$^{*}$Solid State Physics Laboratory, ETH Zurich,
  8093 Zurich, Switzerland \\$^{+}$Angewandte Festk\"{o}rperphysik,
Ruhr-Universit\"{a}t Bochum, 44780 Bochum, Germany}

\begin{abstract}
Strong spin-orbit interaction characteristic for p-type GaAs
systems, makes such systems promising for the realization of
spintronic devices. Here we report on transport measurements in
nanostructures fabricated on p-type, C-doped GaAs heterostructures
by scanning probe oxidation lithography. We observe conductance
quantization in a quantum point contact, as well as pronounced
Coulomb resonances in two quantum dots with different geometries.
Charging energies for both dots, extracted from Coulomb diamond
measurements are in agreement with the lithographic dimensions of
the dots. The absence of excited states in Coulomb diamond
measurements indicates that the dots are in the multi-level
transport regime.

\end{abstract}

\maketitle

%%%%%%%%%%%%%%%%%%%%%%%%%%%%%%%%%%%%%%%%%%%%
%% MAINMATTER
%%%%%%%%%%%%%%%%%%%%%%%%%%%%%%%%%%%%%%%%%%%%

%%\section{Introduction}

The interest in low dimensional hole-doped GaAs systems arises
primarily from the fact that spin-orbit \cite {Winkler03} as well
as carrier-carrier Coulomb interaction effects are more pronounced
in such systems compared to the more established electron doped
systems, due to the fact that holes have approximately 6 times
larger effective mass than electrons \cite {Grbic04}. However, the
investigation of electronic transport in low-dimensional p-type
GaAs systems was mainly limited to two-dimensional bulk samples,
due to difficulties to fabricate stable p-type nanodevices with
conventional split-gate technique. The main problems we
encountered in measurements on split-gate devices tested on
several different p-type heterostructures are strong hysteresis
effects in gate sweeps, as well as significant gate instabilities
and charge fluctuations.

In order to overcome these problems with metallic gates, we employ
a different lithography technique, namely, Atomic Force Microscope
(AFM) oxidation lithography \cite {Held98, Rokhinson02} to define
nanostructures on two-dimensional hole gases (2DHG). We
demonstrate that for a 2DHG 45 nm below the sample surface the AFM
written oxide lines with a height of 15-18 nm completely deplete
the 2DHG beneath at low temperatures \cite {Grbic05}. Density and
mobility of the unpatented sample at 4.2 K are: p =
4$\times$10$^{11}$ cm$^{-2}$, $\mu$=120'000 cm$^2$/Vs.

\begin{figure}
\begin{center}
  \includegraphics{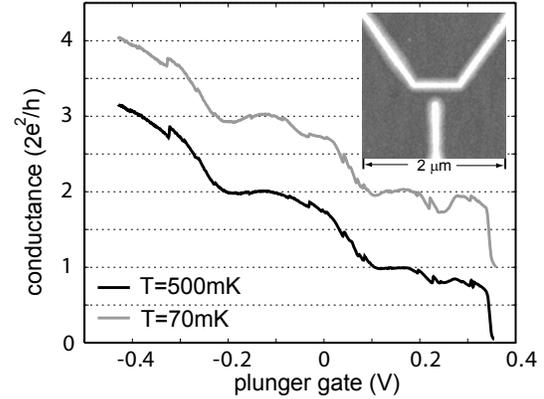}
  \caption{Two-terminal QPC conductance measurement at T=500 mK (black curve) and T=70 mK (gray curve). The trace corresponding to T=70 mK is shifted upwards by one conductance unit for clarity. A bias of 10 $\mu$V is applied symmetrically across the QPC. Inset: AFM micrograph of the
QPC}
  \label{fig1}
  \end{center}
\end{figure}

We fabricated a quantum point contact (QPC) with a lithographic
width of 165 nm and tested its electronical functionality by
measuring its conductance at low temperatures (Fig. \ref{fig1}).
At the temperature of 500 mK quantized conductance plateaus are
observed corresponding to transmission of one and two modes
through the QPC. In addition, a plateau-like structure is observed
at $\sim 0.8\times2e^{2}/h$. As the temperature is reduced to
$\sim70$ mK, this plateau-like feature evolves into a dip-like
structure below the first plateau (Fig. \ref{fig1}). In
differential conductance vs. bias measurements we observe a
pronounced zero-bias peak for a QPC conductance $\sim
0.8\times2e^{2}/h$, which weakens as the conductance increases to
$2e^{2}/h$, and completely disappears above the first plateau.
This behavior might indicate that the structure below the first
plateau is related to Kondo-like effect \cite{Cronenwett02}.
Besides, at T$=70$ mK another plateau-like structure at $\sim
1.7\times2e^{2}/h$ appears. All features observed in this sample
were stable and reproducible in several different cool-downs.

%%\section{Samples}

Hole transport in p-type GaAs quantum dots is also explored. Two
quantum dots were fabricated with AFM lithography - one
rectangular (Fig. 2(b)) with lithographic dimensions
$430\times170$ nm$^2$, and the other circular (Fig. 2(d)) with
lithographic radius $\sim320$ nm. The transport measurements in
both dots have been performed in a dilution refrigerator at a base
temperature of $\sim$ 70 mK. We have measured the two-terminal
conductance of the dots by applying either a small dc or ac bias
voltage V$_{bias}$ between source and drain, and measuring the
current through the dot with a resolution better than 50 fA.

The QPC gates are tuned to configurations where the dots are
symmetrically coupled to the leads. Pronounced Coulomb resonances
are observed in both dots (Fig. 2(a) shows Coulomb peaks from the
rectangular dot). It is important to note that the dots close when
the value of the plunger-gate voltage increases $-$ this is a
clear indication that we measure hole transport. Coulomb
resonances are fitted both with an expression for a thermally
broadened Coulomb blockade peak in the multi-level transport
regime and a coupling broadened Lorentzian peak. In all cases the
thermally broadened resonance fits better to the data than a
coupling broadened resonance, indicating that the dots are in the
weak coupling regime. The electronic temperature extracted from
the fitting is $\sim$ 130 mK.

\begin{figure}
 \begin{center}
  \includegraphics{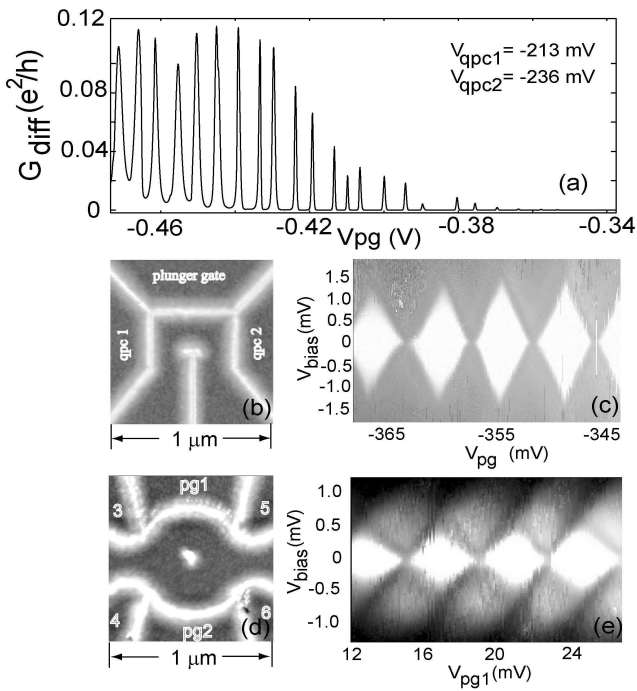}
  \caption{(a) Differential conductance of the rectangular dot (b) in the configuration V$_{qpc1}$ = -213 mV, V$_{qpc2}$ = -236 mV as a function of plunger gate voltage. (b) AFM micrograph of the rectangular quantum dot with designations of the gates. (c) Coulomb diamonds in differential conductance for the rectangular dot in the configuration V$_{qpc1}$ = -225 mV, V$_{qpc2}$ = -235 mV, represented in a logarithmic gray scale plot (white regions represent low conductance). (d) AFM micrograph of the circular quantum dot. (e) Coulomb diamonds in differential conductance for the circular dot in the configuration: V$_{pg2}$ = -32 mV, V$_{3}$ = 72 mV, V$_{4}$ = 120 mV, V$_{5}$ = 310 mV, V$_{6}$ = 200 mV represented in a logarithmic gray scale
  plot.}
  \label{fig2}
  \end{center}
\end{figure}

Coulomb diamond measurements are performed in the weak coupling
regime for both dots, and the results are shown in Fig. 2. The
uniform size of the diamonds indicates that all confined holes
reside in one single potential minimum rather than occupying
several disconnected or tunnel-coupled potential minima. From the
extent of the diamonds in bias direction we estimate a charging
energy of the rectangular dot to be $E_{C,rect}\approx1.5$ meV,
while the lever-arm of the plunger gate is
$\alpha_{rect}\approx0.26$. In case of the circular dot we obtain
$E_{C,circle}\approx0.5$ meV and $\alpha_{circle}\approx0.14$.
Assuming a disk-like shape of the dots allows us to estimate
electronic radius of the dots from the values of their charging
energies. The obtained value for the rectangular dot is
$r_{rect}\approx115$ nm, and for the circular
$r_{circle}\approx340$ nm, which is consistent with the
lithographic dimensions of the dots and indicates that the dots
are really formed in the regions encircled by the oxide lines.

Due to the large effective mass of holes, the single-particle
level spacing in case of hole quantum dots is significantly
smaller compared to electron quantum dots with similar size. The
estimated mean single-particle level spacing in the rectangular
dot is $\triangle_{rect}\leq 15$ $\mu$eV, and in the circular dot
is $\triangle_{circle}\leq 2$ $\mu$eV. Therefore we were not able
to resolve excited states in Coulomb diamond measurements in
neither of the two dots. This fact, together with the observed
temperature dependence of Coulomb peak heights \cite {Grbic05}
indicates that both dots are in the multi-level transport regime.
In order to be able to investigate the single-particle level
spectrum in hole quantum dots, one has to significantly reduce the
lateral dimensions of the dot as well as the hole temperature.

In conclusion, we fabricated tunable nanodevices on p-type GaAs
heterostructures by AFM oxidation lithography. By using this
fabrication technique we were able to overcome the problems with
large hysteresis effects present in gate sweeps in conventional
split-gate defined nanostructures on p-type GaAs, and the
stability of the structures improved as well. Electronic
functionality of these structures was demonstrated by observing
conductance quantization in a QPC, and Coulomb blockade in two
quantum dots with different geometries. Further reduction in size
of the p-type quantum dots is necessary in order to explore the
influence of spin-orbit and carrier-carrier interactions on
single-particle level spectra.

Financial support from the Swiss National Science Foundation is
gratefully acknowledged.

%%%%%%%%%%%%%%%%%%%%%%%%%%%%%%%%%%%%%%%%%%%%%%%%
%% BACKMATTER
%%%%%%%%%%%%%%%%%%%%%%%%%%%%%%%%%%%%%%%%%%%%%%%%

%% If the bibliography is produced without BibTeX comment out the

\bibliographystyle{aipproc}   % if natbib is available

\end{document}